\documentclass[11pt,twoside
]{article}

\usepackage{baaa2011}
\usepackage{graphicx}
\usepackage{subfigure}
\usepackage{psfrag}
\usepackage{amssymb}
\usepackage[spanish,activeacute,english]{babel}
\usepackage[latin1]{inputenc}
\usepackage[T1]{fontenc} 
\usepackage{ae,aecompl} 
\usepackage{latexsym}
\usepackage{verbatim}
\usepackage{amsmath}
\usepackage{stmaryrd}
\usepackage{amsfonts}
\usepackage{amssymb}
\usepackage{wasysym}
\usepackage[colorlinks=true,dvips]{hyperref}

\begin{document}
\myselectenglish
\vskip 1.0cm
\markboth{Torrelles, Patel, Curiel, G\'omez, Anglada, \& Estalella}%
{H$_2$O maser kinematics in massive star-forming regions}

\pagestyle{myheadings}
\vspace*{0.5cm}
\noindent PRESENTACIÓN ORAL
\vskip 0.3cm
\title{Water maser kinematics in massive star-forming regions: Cepheus A and W75N}


\author{J.~M. Torrelles$^{1}$, N.~A. Patel$^2$, S. Curiel$^3$, J.~F. G\'omez$^4$, G. Anglada$^4$, R. Estalella$^5$}

\affil{%
  (1) ICE(CSIC)-UB/IEEC, Barcelona (Spain)\\
  (2) Harvard-Smithsonian, CfA, Cambridge (USA)\\
  (3) IAUNAM, M\'exico D.F. (M\'exico) \\
  (4) IAA(CSIC), Granada (Spain)\\
  (5) UB/IEEC, Barcelona (Spain)\\
}

\begin{abstract} VLBI multi-epoch water maser observations are a powerful tool to study the dense, warm shocked gas very close to massive protostars. The very high-angular resolution of these observations allow us to measure the proper motions of the masers in a few weeks, and together with the radial velocity, to determine their full kinematics. In this paper we present a summary of the main observational results obtained toward the massive star-forming regions of Cepheus A and W75N,  among them: (i) the identification
of different centers of high-mass star formation activity at scales of $\sim$ 100~AU; (ii) the discovery of new phenomena associated with the early stages of high-mass protostellar evolution (e.g., isotropic gas ejections); and (iii) the identification 
of the simultaneous presence of a wide-angle outflow and a highly collimated jet in the massive object Cep A HW2, similar to what is observed in some low-mass protostars. Some of the implications of these results in the study of high-mass star formation are discussed.
\end{abstract}

\begin{resumen}
Las observaciones de VLBI de la emisi\'on m\'aser del vapor de agua son una herramienta muy \'util  para estudiar
los movimientos del gas chocado denso y caliente en las proximidades de las protoestrellas masivas. La alta resoluci\'on angular que se consigue con esas observaciones permite medir los movimientos propios de los m\'aseres en escalas temporales de unas pocas semanas, y junto con la velocidad radial extraer informaci\'on de su cinem\'atica completa. En este art\'{\i}culo presentamos un resumen de los resultados m\'as relevantes obtenidos en las regiones de formaci\'on estelar de Cepheus A y W75N, entre ellos:  (i) la identificaci\'on de m\'ultiples centros de actividad de formaci\'on estelar masiva en escalas espaciales de $\sim$ 100~UA; (ii) el descubrimiento de nuevos fen\'omenos asociados con las primeras etapas de evoluci\'on de las estrellas masivas, tales como las expulsiones de gas isotr\'opicas; y (iii) la identificaci\'on de un flujo molecular de gran \'angulo de apertura junto con un jet muy colimado en el objeto masivo Cep A HW2, parecido a lo que se observa en algunas protoestrellas de baja masa. Discutimos algunas de las implicaciones de esos resultados en el estudio de la formaci\'on estelar masiva.
\end{resumen}

\section{Introduction}

Low-mass protostars are associated with rotating accretion disks (protoplanetary disks). Part of the material is ejected along the polar axis of the disk, in a collimated wind that removes angular momentum and magnetic flux from the system, allowing the accretion to proceed until the star is assembled. In the early stages of evolution of the ``disk-protostar-outflow" systems, arising at scales of $\sim$ 100~AU, the ejected material interacts with the ambient medium, giving rise to phenomena such as Herbig-Haro (HH) emission, molecular outflows, and radio jets (e.g., Anglada 1996, Machida et al. 2008). This accreting scenario seems also valid, as a first approach, for the formation of  stars with masses up to $\sim$ 20~M$_{\odot}$ (e.g., Garay \& Lizano 1999). However, only a few massive 
disk-protostar-outflow systems have been studied in detail at scales of $\leq$ 3000~AU (Patel et al. 2005, Jim\'enez-Serra et al. 2007, Torrelles et al. 2007,  Zapata et al. 2009, Davies et al. 2010, Carrasco-Gonz\'alez et al. 2010a, 2011, Fern\'andez-L\'opez et al. 2011), probably due to strong observational limitations (e.g., small number of massive protostars in comparison with low-mass protostars, and insufficient  angular resolution and sensitivity of the observations to isolate single disk-protostar-outflow systems, given their high degree of clustering and their long distances). 

We find in these observational limitations the main motivations to study massive star-forming regions through water maser emission at $\lambda$ = 1.35~cm, using Very Long Baseline Interferometry (VLBI) techniques. Young massive objects are highly obscured, but the ambient medium is transparent to 
water maser emission at cm wavelengths.  This maser emission is compact ($\leq$ 1 mas), with brightness temperatures exceeding in some cases 10$^{10}$~K, and therefore very useful to study with extremely high angular resolution (better than 1 mas) the motions of the shocked, warm ($\sim$ 500~K), and dense ($\sim$ 10$^8$-10$^9$ cm$^{-3}$) gas in the close proximity (10-1000 AU) of massive protostars (e.g., Reid \& Moran 1981, Imai et al. 2006) In this paper we summarize the main results obtained toward the high-mass star forming regions of Cepheus A and W75N using VLBI multi-epoch water maser observations with an angular resolution of $\sim$ 0.5~mas reported by Torrelles et al. (2001a,b, 2003, 2011). The common general goals of these VLBI observations were to: (i) observe as close as possible the central engine driving the outflow phenomena associated with the massive star formation, (ii) determine the full motions of the masing gas at AU scales, (iii) identify new centers of massive star formation activity in these regions through the presence of maser emission clustering, and (iv) look for new phenomena related to the high-mass star formation processes. 

\section{Cepheus A}

\begin{figure}[!ht]
\begin{center}
 \includegraphics[scale=0.6]{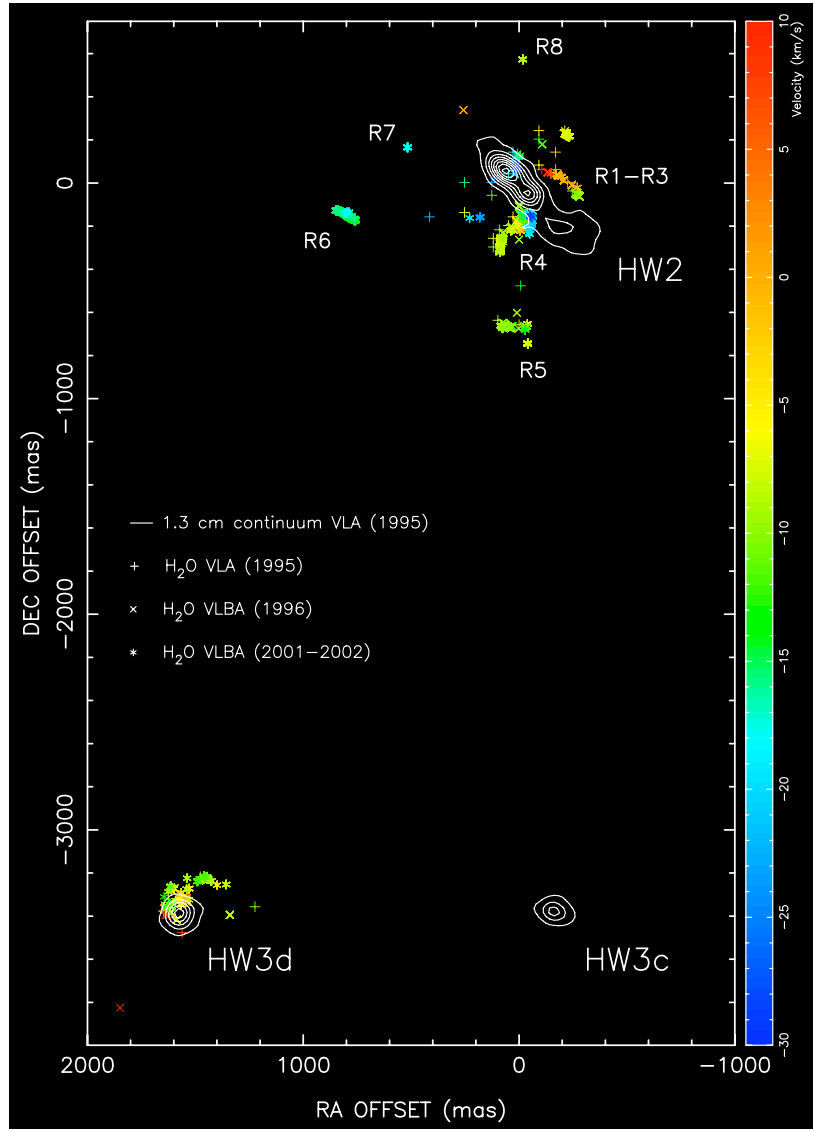}
 \end{center}
 \caption{Positions and radial velocities (colour code) of the water masers overlaid onto the 1.3~cm continuum maps of Cepheus A HW2, HW3c, and HW3d. Sub-regions ``R'' discussed in this paper are numbered. (Figure from Torrelles et al. 2011).}
\end{figure}

At a distance of $\sim$ 700 pc, this is the second nearest high-mass star-forming region after Orion. It contains a cluster of 16 compact ($\sim$ 1$''$) radio continuum sources  (HW objects; Hughes \& Wouterloot 1984, Garay et al. 1996) in a region of $\sim$ 30$''$, with the brightest one (HW2) being a radio jet,
excited by a massive protostar ($\sim$ 15-20~M$_{\odot}$; Rodr\'{\i}guez et al. 1994) that is highly obscured (A$_V$ $\sim$ 10$^3$ mag; Torrelles et al. 1985). What makes HW2 a unique object is its association with a massive disk-protostar-jet system observed at scales of $\sim$ 1000~AU, with a strong magnetic field, oriented parallel to the radio jet moving at $\sim$ 500~km~s$^{-1}$, and powerful water masers (Lada et al. 1981, Patel et al. 2005, Curiel et al. 2006, Jim\'enez-Serra et al. 2007,
Torrelles et al. 2007, 2011, Vlemmings et al. 2010). All these characteristics indicate that the massive HW2 protostar has been formed through an accretion disk, similar to the formation of low-mass stars. 

\begin{figure}[!ht]
\begin{center}
 \includegraphics[scale=0.6]{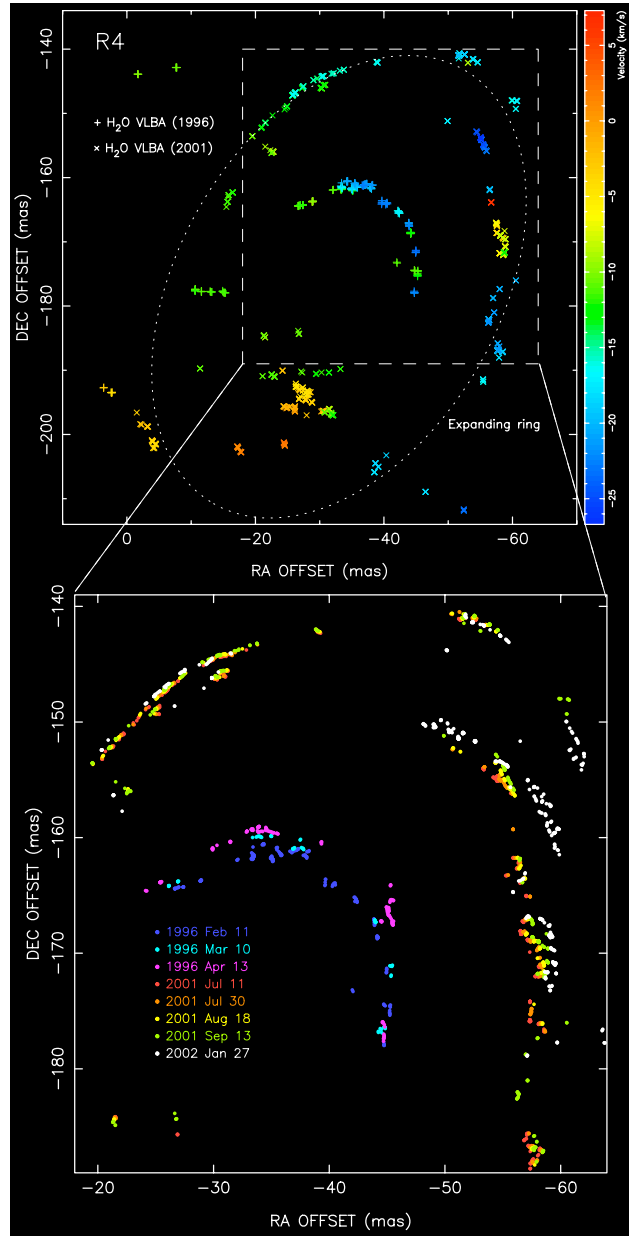} 
 \end{center}
 \caption{{\it Upper panel}: Water maser positions measured in sub-region R4 of Cepheus A (see also Figure~1). In this panel only the 1996 Feb~11 and 2001 Sep~13 VLBA epochs are plotted. {\it Lower panel:} Zoom showing the evolution of the expanding motions in the sky for all the observed VLBA epochs. (Figure from Torrelles et al. 2011).}
\end{figure}

The water maser emission around HW2 has been studied with the Very Large Array (VLA; beam size $\simeq$ 80~mas) and with the Very Long Baseline Array (VLBA; beam size $\sim$ 0.4-0.5 mas)  at nine  different epochs (see Torrelles et al. 2001a,b, 2011 for a detailed presentation of the observations and discussion). In Figure~1, we show the position of the water masers, overlaid onto the 1.3~cm continuum map showing the HW2 radio jet, and the nearby HW3c and HW3d objects (located $\sim$ 3$''$ south from HW2). The VLBA observations revealed that many of the ``single" masers observed with the VLA unfold into remarkable linear/arcuate structures of $\sim$ 40-100~mas size, some of them formed by many linear ``microstructures"  of a few mas in size. The flattened appearance of these water maser linear ``microstructures" and their proper motions indicate that they are originated through shock excitation by winds, as expected from theory (e.g., Elitzur et al. 1992). The different sub-regions where rich structures of masers have been  found around HW2 are labelled as R1-R7 in Figure~1.

In particular, the masers of the R4 sub-region trace a nearly elliptical patchy ring of $\sim$ 70~mas size (50~AU) with expanding motions of $\sim$ 15-30~km~s$^{-1}$ (Figure 2). This expanding ring indicates that it is driven by the wind of a central YSO predicted to be located at a position offset (--0.03$''$, --0.18$''$) from HW2. From the size of the ring, and assuming a constant expansion velocity of $\sim$ 15-30~km~s$^{-1}$, a dynamical age of $\sim$ 4-8~yr is estimated. From Figure 1, we see that the full morphology of R4 is, however,  complex, showing several ``shells", probably produced by multiple ejections in different epochs. The driving source is proposed to be a massive object, to account for the observed very high flux densities of the water masers in this structure ($\sim$ 100~Jy). From the literature, there is no known source at the center of the ring.
Future very high-sensitivity observations at cm and (sub)mm wavelengths (e.g., e-MERLIN, EVLA, SMA) could help to identify this massive object predicted to be located very close (130~AU) to the HW2 massive protostar.

Another subregion with strong water masers showing high-mass star formation activity is R5, located $\sim$ 0.6$''$ south of HW2. In the 1996 observations, R5 contained a remarkable arc structure of $\sim$ 100~mas size defining a circle of radius $\sim$ 90~mas (60~AU) with an accuracy of $\sim$ 1/1000.
This structure was expanding at $\sim$ 9~km~s$^{-1}$, and was interpreted as caused by a short-lived episodic spherical wind (dynamical age $\sim$ 30 yr) from a massive object located at the center of the structure. In fact, after the discovery of this water maser structure, Curiel et al. (2002) detected a weak radio continuum source at that position. This maser structure, five years after its first detection, underwent a distortion in its expansion through the circumstellar medium, losing its degree of symmetry. The importance of this result resides in the fact that isotropic ejections of material from YSOs are  not predicted by current theories of star formation, since the launching mechanism of mass-loss is believed to be the transformation of rotational energy of the disk into collimated outflows via magnetohydrodynamics (MHD) mechanisms. Since the recognition of this isotropic ejection in R5, other cases have been reported in the literature, with water maser structures
tracing expanding motions in multiple directions with respect to a central massive object (e.g., W75N, see \S 3). These results indicate that non-collimated, episodic ejections of material may occur at the earliest stages of evolution of massive objects. Different scenarios have been considered to explain these phenomena, including that the expanding motions of these isotropic outflows are due to the radiation pressure of a central YSO, or that
the water masers are produced in a shocked layer of ambient material around a very young expanding HII region, but both scenarios
are considered to be very unlikely (see Torrelles et al. 2003).

\begin{figure}[!ht]
\begin{center}
 \includegraphics[scale=.44]{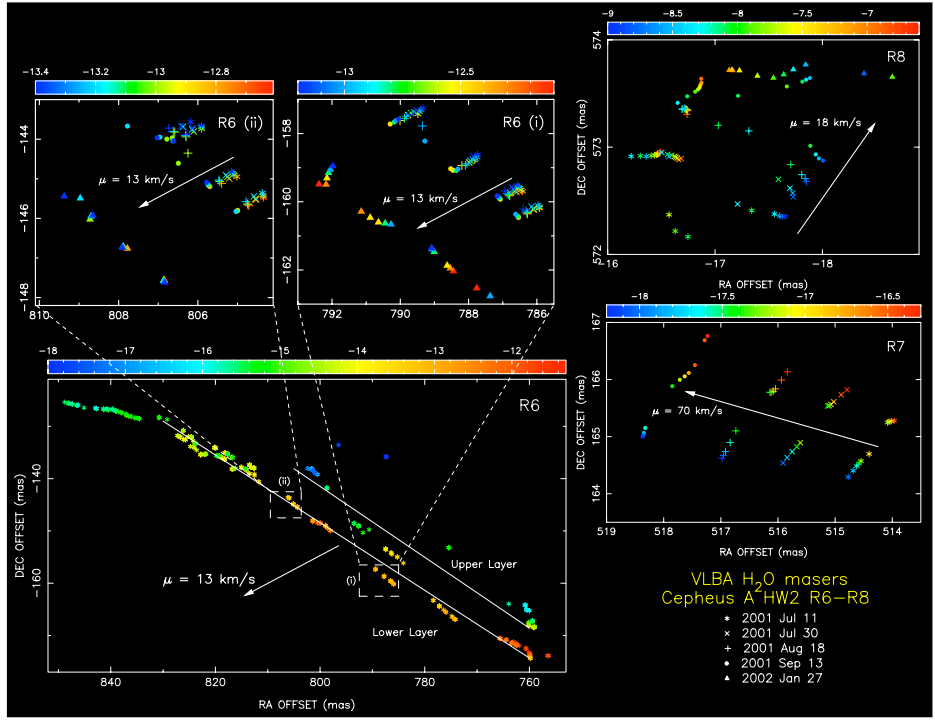}
 \end{center}
 \caption{Position and proper motions of the water masers measured
in sub-regions R6, R7, and R8 of Cepheus A (see also Figure 1) trough five different epochs. Colour code indicates the LSR radial velocity (km~s$^{-1}$) of the masers. (Figure from Torrelles et al. 2011).}
 \end{figure}

The remaining masers of the subregions R1, R2, R3, R6, R7, and R8 (see Figure 1) are interpreted as tracing gas excited  by a wide-angle outflow from the HW2 disk-protostar system. This could explain the fact that the R6, R7, and R8 masers are located near the northeastern edge of the HW2 dust/molecular disk observed by Patel et al. (2005), and present expanding motions covering a wide range of directions. In fact, while R6 is moving at $\sim$ 13~km~s$^{-1}$ toward the southeast with a difference in PA of $\simeq$ 73$^{\circ}$ with respect to the direction of the HW2 radio jet, R8 is moving at $\sim$ 18~km~s$^{-1}$ toward the northwest with a difference in PA of $\simeq$ --80$^{\circ}$. On the other hand, the R7 masers (located in between R6 and R8) have faster motions ($\sim$ 70~km~s$^{-1}$), moving  toward the northeast, along a direction closer to the HW2 radio jet direction, but still with a significant difference in position angle of $\sim$ 30$^{\circ}$ with respect to the motions of the radio jet (see Figure 3). These characteristics can be explained in a scenario where R6 and R8 are shock fronts from the walls of expanding cavities of the circumstellar gas created by a wide-angle wind from HW2 with outflow opening angle of $\sim$ 100$^{\circ}$. In this scenario, the R1-R3 masers located in the opposite side, moving at $\sim$ 5~km~s$^{-1}$ and with a difference in PA of  $\simeq$ --80$^{\circ}$ with respect to the radio jet direction, would represent shocked walls of the southwestern cavities (see Figure 4).

\begin{figure}[!ht]
\begin{center}
 \includegraphics[scale=0.6]{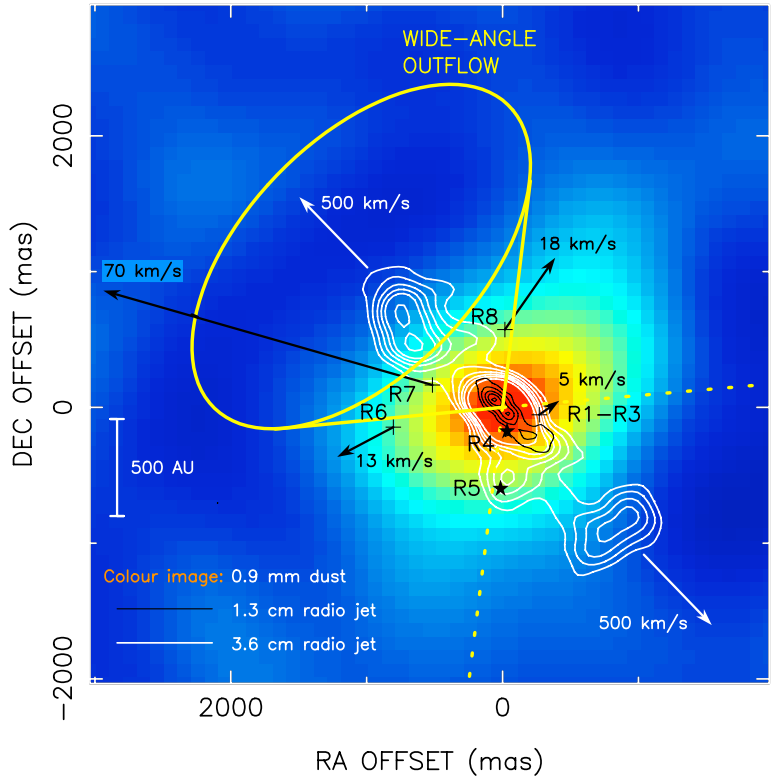}
\end{center} 
 \caption{Wide-angle outflow and jet in Cepheus A HW2. The radio jet (opening angle of $\sim$ 18$^{\circ}$) exhibits ejections in opposite directions, moving away at $\sim$ 500~km~s$^{-1}$ from the central source, which is surrounded by a dust/molecular disk (Curiel et al. 2002, Patel et al. 2005). R6, R8, and R1-3 trace emission fronts from the shocked walls of expanding cavities, created by the wide-angle wind of HW2 (opening angle of $\sim$ 102$^{\circ}$). The R7 masers, with motions along an axis at an angle
 of $\sim$ 30$^{\circ}$ with respect to the radio jet axis,  are excited inside the cavity by the wide-angle wind. They exhibit higher velocity than R6, R8, and R1-3 (which are located at the expanding cavity walls) but lower than the velocity of the jet. The R6, R7, and R8 masers (observed towards the blue-shifted lobe of the 1 arcmin, large-scale bipolar molecular outflow; G\'omez et al. 1999) are blue-shifted with respect to the systemic velocity of the circumstellar disk, while R1-3 (observed towards the red-shifted lobe of the large-scale molecular outflow) are red-shifted. The position of the two massive YSOs required to excite the R4 and R5 maser structures are indicated by star symbols (see text). The star associated with R4 is not yet detected. (Figure adapted from Torrelles et al. 2011).}
\end{figure}

These results imply the simultaneous presence of a wide-angle outflow and a highly collimated jet at a similar physical scale in a massive object, as seen previously in low-mass YSOs (e.g., see Velusamy et al. 2011 and references therein). Different theoretical models have been proposed to explain these two kind of outflows in low-mass YSOs (e.g., ``X-wind", ``Disk-wind" models; see  Machida et al. 2008), but unfortunately, current observations do not have enough angular resolution and sensitivity to test them. What we are observing now in the massive object HW2 is analogous to what is observed in low-mass objects, further supporting that massive stars form in a similar way as low-mass stars. In addition, these observations on HW2 should provide important observational constraints for future models trying to reproduce the presence of outflows with different opening angles in high-mass YSOs. 

\section{W75N}

\begin{figure}[!ht]
\begin{center}
 \includegraphics[scale=.58]{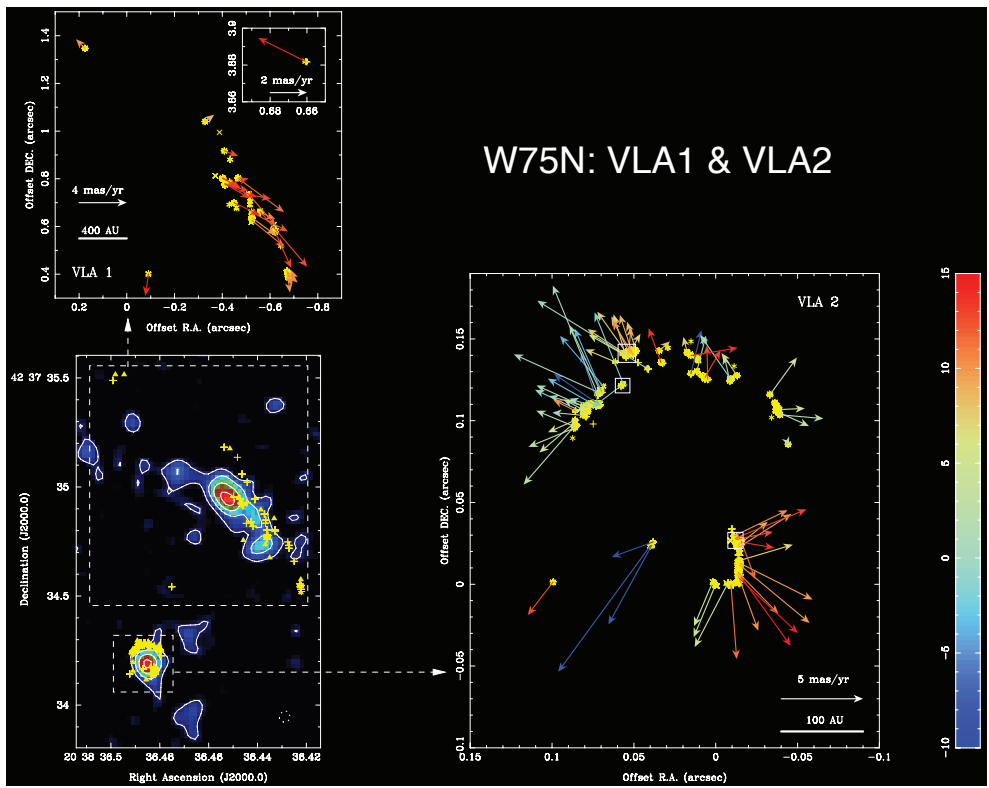}
 \end{center}
 \caption{{\it Bottom left:} Contour map of the 1.3~cm continuum emission of VLA 1 and 2 in W75N.
The positions of the H$_2$O masers detected with the VLA (triangles) and VLBA (plus symbols) are indicated. {\it Top left and bottom right}: Proper motions of the masers measured with the VLBA in VLA 1 and VLA 2. Color 
code indicates the LSR velocity of the masers in km~s$^{-1}$. (Figure from Torrelles et al. 2003).}
 \end{figure}

This massive star-forming region contains a large-scale bipolar molecular outflow,  mid-infrared,  millimeter, and centimeter continuum sources, as well as several maser species (see, e.g.,  Persi et al. 2006 and Carrasco-Gonz\'alez et al. 2010b for a review of the main characteristics of this region). Three of the YSOs (VLA 1, VLA 2, and VLA 3, probably excited by early B-type stars) are located within a region of $\sim$ 1.5$''$ (3000 AU at a distance of 2 kpc) and have been proposed to be at different evolutionary stages. In particular, while in VLA 1 the radio continuum emission and the water masers observed with the VLBA trace a collimated jet at a 2000 AU scale, in VLA 2 the radio continuum source is compact and the water masers move outward in multiple directions at $\sim$ 30~km~s$^{-1}$, tracing a shell-like outflow at a 160 AU scale with a dynamical age of $\sim$ 13~yr (Figure 5). Since both sources share the same molecular gas environment (they are $\sim$ 1400 AU apart in the sky), Torrelles et al. (2003) proposed that the very different degree of outflow collimation in these two sources is not only a consequence of ambient conditions, but probably due to the different evolutionary states of the individual YSOs, with VLA 2 in an earlier stage of evolution than VLA 1 (the VLA 2 non-collimated outflow is observed at a smaller scale in comparison with the highly collimated outflow in VLA 1, which probably excites the large-scale bipolar molecular outflow in the region). These authors also suggest that the VLA 2 outflow could evolve in the future into a collimated jet as the one observed in VLA 1, and in fact, more recent VLBA water maser observations carried out by Surcis et al. (2011) have provided some indications that the formation of a jet might be taking place in VLA 2.  

In summary, the VLBA water maser observations in W75N have shown a well differentiated outflow geometry in two close similar YSOs, providing another example of a $\sim$ 100 AU scale non-collimated outflow associated with the early stages of evolution of massive objects. As in the case of R5 and R4 in Cepheus A (\S~2), it is still unclear what is the physical mechanism that produces these kind of ``short-lived" non-collimated outflows, opening new challenges to advance in the knowledge of the earliest stages of stellar evolution.

\section{Conclusions}

VLBI multi-epoch water maser observations are a very powerful tool to study the gas very close to the central engine driving the outflow phenomena associated with massive star formation. With these observations it has been possible to identify new centers of high-mass star formation at scales of $\sim$ 100~AU. In addition, these observations are revealing unexpected phenomena in the earliest stages of  evolution of massive objects (e.g., non-collimated ``short-lived" outflows in different massive YSOs), and providing new insights in the study of the dynamic scenario of the formation of high-mass stars (e.g., simultaneous presence of a jet and wide-angle outflow in the massive object Cep A HW2, similar to what is observed  in low-mass YSOs).

GA, RE, JFG, and JMT acknowledge support from MICINN AYA2008-06189-C03 grant co-funded with FEDER funds.

\begin{referencias}
\reference Anglada, G. 1996, ASPC, 93, 3
\reference Carrasco-Gonz\'alez, C., Rodr\'{\i}guez, L. F.,  Anglada, G.,  Mart\'{\i}, J., Torrelles, J. M., Osorio, M. 2010a, Science, 330, 1209
\reference Carrasco-Gonz\'alez, C., Rodr\'{\i}guez, L. F., Osorio, M., Anglada, G., Mart\'{\i}, J., Torrelles, J. M., Galv\'an-Madrid, R., D'Alessio, P. 2011, RMxAC, 40, 229 
\reference Carrasco-Gonz\'alez, C., Rodr\'{\i}guez, L. F., Torrelles, J. M., Anglada, G., Gonz\'alez-Mart\'{\i}n, O. 2010b, AJ, 139, 2433
\reference Curiel, S., Ho, P. T. P., Patel, N. A.,  
Torrelles, J. M., Rodr\'{\i}guez, L. F., Trinidad, M. A., 
Cant\'o, J., Hern\'andez, L., G\'omez, J. F.,  Garay, G.,  
Anglada, G. 2006, ApJ, 638, 878
\reference Curiel, S., Trinidad, M. A., Cant\'o, J., 
Rodr\'{\i}guez, L. F., Torrelles, J. M., Ho, P. T. P., 
Patel, N., Greenhill, L., G\'omez, J. F., Garay, G., Hern\'andez, L., 
Contreras, M. E., Anglada, G. 2002, ApJ, 564, L35
\reference Davies, B., Lumsden, S. L., Hoare, M. G., Oudmaijer, R. D., de Wit, W-J. 2010, MNRAS, 402, 1504
\reference Elitzur, M., Hollenbach, D. J., McKee, C. F. 1992, ApJ, 394, 221
\reference Fern\'andez-L\'opez, M., Girart, J. M., Curiel, S., G\'omez, Y., Ho, P. T. P., Patel, N. 2011, AJ, 142, 97
\reference Garay, G., Ram\'{\i}rez, S., Rodr\'{\i}guez, L.~F., Curiel, 
S., Torrelles, J.~M. 1996, 1996, ApJ, 459, 193
\reference Garay, G., Lizano, S. 1999, PASP, 111, 1049
\reference G\'omez, J. F., Sargent, A., Torrelles, J. M.,
Ho, P. T. P., Rodr\'{\i}guez, L. F., Cant\'o, J., Garay, G. 1999, ApJ, 514, 287
\reference Hughes, V. A., Wouterloot, J. G. A. 1984, ApJ, 276, 204
\reference Imai, H. et al.  2006, PASJ, 58, 883
\reference Jim\'enez-Serra, I. et al. 2007, ApJ, 661, L187
\reference Lada, C. J., Blitz, L., Reid, M. J.,  Moran, J. M. 1981, ApJ, 243, 769
\reference Machida, M. N., Inutsuka, S.-i., Matsumoto, T. 2008, ApJ, 676, 1088
\reference Patel, N. A., Curiel, S., Sridharan, T. K., 
Zhang, Q., Hunter, T. R., Ho, P.T.P., Torrelles, J. M., Moran,
J. M., G\'omez, J. F. G., Anglada, G. 2005, Nature, 437, 109 
\reference Persi, P., Tapia, M., Smith, H. A. 2006, A\&A, 445, 971
\reference Reid, M. J., Moran, J. M. 1981, ARAA, 19, 231
\reference Rodr\'{\i}guez, L.~F., Garay, G., Curiel, S.,
Ram\'{\i}rez, S., Torrelles, J.~M., G\'{o}mez, Y., Vel\'{a}zquez, A. 1994, ApJ, 430, L65
\reference Surcis, G., Vlemmings, W. H. T., Curiel, S., Hutawarakorn Kramer, B., Torrelles, J. M., Sarma, A. P. 2011, A\&A, 527, 48
\reference Torrelles, J. M., Ho, P. T. P., Rodr\'{\i}guez, L. F.,  Cant\'o, J. 1985, ApJ, 288, 595
\reference Torrelles, J. M., Patel, N., Anglada
G., G\'omez, J. F., Ho, P. T. P., Cant\'o, J., Curiel, S., 
Lara, L., Alberdi, A., Garay, G., Rodr\'{\i}guez, L. F. 2003, ApJ, 598, L115
\reference Torrelles, J. M., Patel, N. A., Curiel, S., Estalella, R., G\'omez, J. F., Rodr\'{\i}guez, L. F., Cant\'o, J., Anglada, G., Vlemmings, W., Garay, G., Raga, A. C., Ho, P. T. P. 2011, MNRAS, 410, 627
\reference Torrelles, J. M., Patel, N. A., Curiel, S., Ho, P. T. P., Garay, G., Rodr\'{\i}guez, L. F. 2007, ApJ, 666, L37
\reference Torrelles, J. M., Patel, N., G\'omez, J. F., 
Ho, P. T. P., Rodr\'{\i}guez, L. F., Anglada, G., Garay, G., Greenhill, L.,
Curiel, S., Cant\'o, J. 2001a, Nature, 411, 277
\reference Torrelles, J. M., Patel, N., G\'omez, J. F., 
Ho, P. T. P., Rodr\'{\i}guez, L. F., Anglada, G., Garay, G., Greenhill, L.,
Curiel, S., Cant\'o, J. 2001b, ApJ, 560, 853
\reference Velusamy, T., Langer, W. D., Kumar, M. S. N., Grave, J. M. C. 2011, ApJ, 741, 60
\reference Vlemmings, W. H. T. et al. 2010, MNRAS, 404, 134
\reference Zapata, L. A., Ho 
P. T. P., Schilke, P., Rodr{\'{\i}}guez, L. F., Menten, K., Palau, A., 
Garrod, R. T.  2009, ApJ, 698, 1422 
\end{referencias}
\end{document}